\documentclass[aps,floats]{revtex4}
\usepackage{amsmath,amssymb}
\usepackage{graphicx,epsfig}
\begin{document}
\bibliographystyle {plain}

\def\oppropto{\mathop{\propto}} 
\def\opsimeq{\mathop{\simeq}}
\def\opoverderline{\mathop{\overline}}
\def\operarrow{\mathop{\longrightarrow}}
\def\opsim{\mathop{\sim}} 
\def\oplim{\mathop{\lim}} 

\def\fig#1#2{\includegraphics[height=#1]{#2}}
\def\figx#1#2{\includegraphics[width=#1]{#2}}

%\newcommand{\fig}[2]{\epsfxsize=#1\epsfbox{#2}} \reversemarginpar 

%%%%%%%%%%%%%%%%%%%%%%%%%%%%%%%%%%%%%%%%%%%%%%%%%%%%%%%%%%%%%%%%%%%%%%%%%%%%
\title{ Random Transverse Field Ising Model in dimension $d>1$ : \\
scaling analysis in the disordered phase from the Directed Polymer model } 

%%%%%%%%%%%%%%%%%%%%%%%%%%%%%%%%%%%%%%%%%%%%%%%%%%%%%%%%%%%%%%%%%%%%%%%%%%%%

\author{ C\'ecile Monthus and Thomas Garel }
 \affiliation{Institut de Physique Th\'{e}orique, CNRS and CEA Saclay
91191 Gif-sur-Yvette cedex, France}

\begin{abstract}

For the quantum Ising model with ferromagnetic random couplings $J_{i,j}>0$ and random transverse fields $h_i>0$ at zero temperature in finite dimensions $d>1$, we consider the lowest-order contributions in perturbation theory in $(J_{i,j}/h_i)$ to obtain some information on the statistics of various observables in the disordered phase. We find that the two-point correlation scales as :  $\ln C(r) \sim - \frac{r}{\xi_{typ}} +r^{\omega} u$, where $\xi_{typ} $ is the typical correlation length, $u$ is a random variable, and $\omega$ coincides with the droplet exponent $\omega_{DP}(D=d-1)$ of the Directed Polymer with $D=(d-1)$ transverse directions. Our main conclusions are (i) whenever $\omega>0$, the quantum model is governed by an Infinite-Disorder fixed point : there are two distinct correlation length exponents related by $\nu_{typ}=(1-\omega)\nu_{av}$ ; the distribution of the local susceptibility $\chi_{loc}$ presents the power-law tail $P(\chi_{loc}) \sim 1/\chi_{loc}^{1+\mu}$  where $\mu$ vanishes as $\xi_{av}^{-\omega} $, so that the averaged local susceptibility diverges in a finite neighborhood $0<\mu<1$ before criticality (Griffiths phase) ; the dynamical exponent $z$ diverges near criticality as $z=d/\mu \sim \xi_{av}^{\omega}$ (ii) in dimensions $d \leq 3$, any infinitesimal disorder flows towards this Infinite-Disorder fixed point with $\omega(d)>0$ (for instance $\omega(d=2)=1/3$ and $\omega(d=3) \sim 0.24$) (iii) in finite dimensions $d > 3$, a finite disorder strength is necessary to flow towards the Infinite-Disorder fixed point with $\omega(d)>0$ (for instance $\omega(d=4) \simeq 0.19$), whereas a Finite-Disorder fixed point remains possible for a small enough disorder strength. For the Cayley tree of effective dimension $d=\infty$ where $\omega=0$, we discuss the similarities and differences with the case of finite dimensions.

\end{abstract}

\maketitle

\section{ Introduction }

The quantum Ising model  
\begin{eqnarray}
{\cal H}_{pure} =  - J \sum_{<i,j>} \sigma^z_i \sigma^z_j - h \sum_i \sigma^x_i
\label{hpure}
\end{eqnarray}
involving a ferromagnetic coupling $J>0$ between nearest neighbors $<i,j>$,
and a transverse field $h$ on each site,
is the basic model to study quantum phase transitions at zero-temperature \cite{sachdev}.
This model on an hypercubic lattice in dimension $d$ is well understood
via the equivalence with the classical Ising model in dimension $d^{class}=d+1$,
i.e. the time plays the role of an extra space-dimension \cite{sachdev},
and the dynamical exponent is $z_{pure} = 1$.
The lower-critical dimension $d^{class}_l=1$ and the upper critical dimension $d^{class}_u=4$
for the classical model yields $d_l=0$ and $d_u=3$ for the quantum model.  
Above the upper critical dimension $d \geq d_u=3$, the mean-field critical exponents
($\alpha_{MF} =0 ; \beta_{MF} = 1/2 ; \gamma_{MF} = 1 ; \nu_{MF} = \frac{1}{2}$) are exact. 
In dimension $d=1$, the critical exponents are exactly known from the $d^{class}=2$ classical Ising model.

Let us now consider the disordered model, where the nearest-neighbor couplings $J_{i,j}>0$
and the transverse-fields $h_i>0$ are independent random variables
drawn with two distributions $\pi_{coupling}(J)$ and $\pi_{field}(h)$
\begin{eqnarray}
{\cal H} =  -  \sum_{<i,j>} J_{i,j}  \sigma^z_i \sigma^z_j - \sum_i h_i \sigma^x_i
\label{hdes}
\end{eqnarray}
In dimension $d=1$, exact results for a large number of observables
have been obtained by Daniel Fisher \cite{fisher} 
 via the asymptotically exact strong disorder renormalization procedure
 (for a review, see \cite{review_strong}). In particular, the transition is governed
by an Infinite-Disorder fixed point and presents unconventional scaling laws with respect to the pure case.
In dimension $d>1$, the strong disorder renormalization procedure can still be defined.
It cannot be solved analytically, because the topology of the lattice changes upon renormalization,
but it has been studied numerically with the conclusion that the transition is also governed by
an Infinite-Disorder fixed point in dimensions $d=2,3,4$ 
 \cite{motrunich,fisherreview,lin,karevski,lin07,yu,kovacsstrip,kovacs2d,kovacs3d,kovacsentropy,kovacsreview}.
These numerical renormalization results are in agreement with the results of independent quantum Monte-Carlo
in $d=2$ \cite{pich,rieger}.

In dimension $d=1$, one has the important additional property that any infinitesimal disorder 
flows towards the Infinite-Disorder fixed point \cite{fisher,review_strong}.
To try to answer this question in dimension $d>1$, the first natural step is to determine 
the relevance of disorder at the transition of the pure model
via the use of the Harris criterion \cite{harris}  or the inequality $\nu \geq 2/d$ of Chayes {\it et al} \cite{chayes}.
Below the upper-critical dimension $d \leq d_u=3$ of the pure quantum model, 
there exists a single correlation exponent $\nu_{pure}$,
and the conclusion is that disorder is relevant for $\nu_{pure}<2/d$, and irrelevant for $\nu_{pure}>2/d$.
Here, the conclusion is thus that disorder is relevant in dimension $d=1$ where $\nu_{pure}(d=1)=\nu_{pure}^{class}(d^{class}=2)= 1 <2/d=2$, in dimension $d=2$ where $\nu_{pure}(d=2)=\nu_{pure}^{class}(d^{class}=3) \simeq 0.63.. <2/d =1$, and in dimension $d=3$ where $\nu_{pure}(d=3)=\nu_{pure}^{class}(d^{class}=4)= 1/2  <2/d =2/3$. Above the upper-critical dimension $d \geq d_u=3$ of the pure model, there exists two correlation length
exponents : the usual one $\nu^{MF}=1/2$ defined from the two-point correlation,
and the exponent $\nu_1$ that governs finite-size scaling,
where $\nu_1^{class}=2/d^{class}$ for the classical Ising model in dimension $ d^{class}\geq d^{class}_u=4$  \cite{fssabovedc},
and where $\nu_1^{quantum}=3/(2d)$ for the quantum Ising model in dimension $d \geq d_u=3$ \cite{nishiyama}
(because the quantum model in dimension $d$ is equivalent to a $(d+1)$ classical model in a special cylindrical geometry,
see \cite{nishiyama} and references therein).
So the use of the Harris criterion is more involved, since the two values $\nu$ and $\nu_1$
can give different answers \cite{sarlat}.
In our present case, the Harris criterion based on $\nu=1/2$ yields that disorder should be irrelevant 
for $d \geq 4$ \cite{fisherreview,motrunich,sachdev}, whereas the Harris criterion based on $\nu_1=3/(2d)$
leads to the conclusion that disorder should be always relevant !
In addition, these arguments based on the weak-disorder Harris criterion have been questioned \cite{motrunich,fisherreview} since rare regions can play an essential role in these quantum disordered models, where the disorder is
actually 'infinitely' correlated along the time-direction. The effects of rare regions are discussed in detail in the
review \cite{vojta}, with the conclusion that the important parameter is the effective dimensionality $d_{RR}$
of rare regions (in our present quantum model where the disorder is actually 'infinitely' 
correlated along the time-direction, the dimensionality of rare regions is $d_{RR}=1$ \cite{vojta}).
This dimensionality $d_{RR}$ of rare regions should be compared to the lower critical dimension $d^{class}_l$
sufficient to obtain ordering  (in our present quantum model where the order is a ferromagnetic magnetization,
the lower critical dimension is $d^{class}_l=1$). The random quantum Ising model thus corresponds to 
the case $d_{RR}=d^{class}_l$, i.e. to the so-called class B in the classification described in \cite{vojta},
 where rare regions dominate the critical behavior and induce an unusual activated scaling.
(The conventional power-law scaling is expected
 to hold for the so-called class A corresponding to $d_{RR}<d^{class}_l$, 
where rare regions cannot undergo the phase transition by themselves, 
whereas the so-called class C corresponds to the case $d_{RR}>d^{class}_l$
 where rare regions can order by themselves
at different values of the order parameter). 
In conclusion, this argument based on rare regions suggests that the unusual activated scaling
found by Daniel Fisher \cite{fisher} in dimension $d=1$ via 
the asymptotically exact strong disorder renormalization procedure 
should  exist in any finite dimension $d$ and even on the Cayley tree ($d=\infty$).

To elucidate what happens in high finite dimension $d$, it is natural to consider
 the limit of infinite dimension $d=\infty$. 
Two opposite conclusions have appeared recently :
on one hand, an Infinite-Disorder fixed point has  been found numerically 
for the Erd\"os-R\'enyi random graph in \cite{kovacs3d}, but on the other hand,
the solution on the Cayley tree obtained via some approximations within 
the quantum cavity framework  \cite{ioffe,feigelman,dimitrova}
has been interpreted as a 'Finite-Disorder critical point'
 (with however strong inhomogeneities).

In this paper, our aim is to clarify the possibility of an 'Infinite Disorder' 
fixed point as a function of the dimension $d$ via simple scaling arguments.
We focus on the {\it disordered phase } of the random quantum Ising model  Eq. \ref{hdes},
and  we consider the leading contributions in perturbation theory in 
the disordered variables $J_{i,j}/h_i$ for various observables. 
We conclude that the properties of the quantum model in its disordered phase 
are determined by the problem of the Directed Polymer in a random medium 
with $D=(d-1)$ transverse directions.
We analyze the various consequences of this correspondence.
Since the solution on the Cayley tree obtained via the approximated quantum cavity method 
 \cite{ioffe,feigelman,dimitrova} 
 is also related to the problem of the Directed Polymer on
a tree, we rediscuss this case to emphasize the similarities and differences between
 finite dimensions and infinite dimension.

The paper is organized as follows.
In section \ref{sec_perturb}, we compute the magnetization of a center-site when an infinitesimal magnetization $m_{ext} \to 0$ is imposed on the surface of a finite sample, via a
 perturbative expansion deep in the disordered phase, and
we explain the link with the scaling properties of the Directed Polymer model. In section \ref{sec_corre}, we analyze the consequences for the two-point correlation, and obtain that there are two distinct correlation length exponents  $\nu_{typ}$ and $\nu_{av}$ for the typical and averaged correlation lengths.
In section \ref{sec_chi}, we discuss the power-law tail of the distribution 
$P(\chi_{loc}) \sim 1/\chi_{loc}^{1+\mu}$ of 
  the local susceptibility $\chi_{loc}$, where $\mu$ 
vanishes near criticality.
In section \ref{sec_dyn}, we analyze the statistics of the gap and find that
the dynamical exponent diverges near criticality as $z=d/\mu$.
Finally in section \ref{sec_tree}, we consider the case of the Cayley tree 
to compare with the case of finite dimensions $d$.
Section \ref{sec_conclusion} summarizes our conclusions.
 Appendix \ref{app_DP} contains a reminder on properties of the Directed Polymer model
that are used in the text.
In Appendix \ref{app_anderson}, we discuss the analogy with Anderson localization, where
the droplet exponent of the Directed Polymer also appears in the localized phase.

\section{ Magnetization of a center site }

\label{sec_perturb}

\subsection{ Perturbative expansion deep in the disordered phase }

When the ferromagnetic couplings vanish $J_{i,j} =0$, all spins are completely independent,
there is no order even at the distance of one lattice spacing.
Deep in the disordered phase of the quantum model, where the correlation length remains short,
it is thus natural to consider this uncorrelated state as the reference state, 
and to study a perturbative expansion in the couplings $J_{i,j}$.

In this section, we consider the magnetization $m_0=<\sigma^z (\vec 0) >$
of a center site $\vec 0$ in the ground state of a finite sample of volume $(2L)^d$
where an infinitesimal magnetization $m_{ext} \to 0$ is imposed on the surface $S$ of the sample.
At first order in the couplings $J_{i,j}$, the magnetization  $m_0$
is related to the magnetizations $m_j$ of its $K=(2d)$ neighbors on the hypercubic lattice
\begin{eqnarray}
m_0 = \sum_{j=1}^{2d} \frac{J_{0,j}}{h_0} m_j
\label{mperturbfirst}
\end{eqnarray}
If we iterate up to the boundaries to reach the imposed infinitesimal magnetization $m_{ext} \to 0$,
we obtain a sum over the paths $P$ going from the center site $\vec 0$ to the surface $S$
\begin{eqnarray}
m_0 = m_{ext}  \sum_{ Path P: {\vec 0} \to S} W(P)
\label{mperturb}
\end{eqnarray}
where the weight $W(P) $ of each path $P$ is the product of the factors 
$\left(\frac{J_{i,j}}{h_i} \right)$ along the path
\begin{eqnarray}
W(P) = \left( \prod_{(i,j) \in P} \frac{J_{i,j}}{h_i} \right)
\label{wP}
\end{eqnarray}

\subsection{ Equivalence with the Directed Polymer model}

In the whole disordered phase, the magnetization $m_{ 0}$ of the center site
should typically decay exponentially with respect to the distance $L$ to the boundary
\begin{eqnarray}
\ln \frac{m_0} {m_{ext}}  \opsimeq_{ L \to +\infty}  - \frac{L }{\xi_{typ}} +...
\label{defxityp}
\end{eqnarray}
 where $\xi_{typ} $ represents the typical correlation length of the disordered phase.

Deep in the disordered phase where we have written the perturbative expansion of Eq. \ref{mperturb},
we thus expect that the weight $W(P)$ of each path $P$ will typically decay exponentially
with the length $l(P)$ of the path.
As a consequence, the sum over paths in Eq. \ref{mperturb} will be dominated by
{\it directed paths} going from the center site to the surface at distance $L$.
 So the scaling properties will be the same as for the Directed Polymer
in a random medium with $D=(d-1)$ transverse directions
(see Appendix \ref{app_DP} for a reminder on the Directed Polymer model).
In particular, whenever the Directed Polymer is in its localized phase 
characterized by the droplet exponent $\omega=\omega_{DP}(D)$ (see section \ref{app_DPloc} in Appendix),
the center-site magnetization will presents the scaling
\begin{eqnarray}
\ln \frac{m_0} {m_{ext}}  \opsimeq_{ L \to +\infty} 
 - \frac{L }{\xi_{typ}} +  L^{\omega} u
\label{xitypomega}
\end{eqnarray}
where $u$ is a random variable of order $O(1)$.

Deep in the disordered phase, the mapping to the Directed Polymer model should 
be quantitative,
i.e. $m_0$ of Eq. \ref{mperturb} corresponds to the partition function of Eq. \ref{directed}
at temperature $T=1$ with effective random energies $\epsilon_s( i )$
on sites $i$ determined by the transverse fields $h_i$
\begin{eqnarray}
\epsilon_s( i ) = \ln h (i)
\label{epssite}
\end{eqnarray}
and with effective random energies $\epsilon_b( i,j )$
on bonds determined by the ferromagnetic couplings $J_{i,j}$
\begin{eqnarray}
\epsilon_b( i,j ) = - \ln J_{i,j}
\label{epsbond}
\end{eqnarray}
From the dimensionality $D=d-1$ and the disorder distribution of this effective Directed Polymer model,
one can determine whether the Directed Polymer is in its localized phase or not.
From the reminder of section \ref{app_phasediagram}, we conclude~:

(i)  In dimension $d \leq 3$ for the quantum model, there is no delocalized phase 
for the associated Directed Polymer model,
so that any initial disorder should correspond
 to the localized phase of the Directed Polymer
characterized by the droplet exponent $\omega(d)=\omega_{DP}(D=d-1)$.
In particular, for the quantum model in dimension $d=2$, we conclude 
that the corresponding exponent 
should take the explicit value (Eq. \ref{omegaD1})
\begin{eqnarray}
 \omega(d=2)=\omega_{DP}(D=1)=\frac{1}{3}
\label{omegaD1texte}
\end{eqnarray}
whereas in $d=3$, it should be of order $\omega(d=3)=\omega_{DP}(D=2)\simeq 0.24 $
(Eq. \ref{omeganumeD}).

(ii) In dimension $d>3$ for the quantum model, 
the associated Directed Polymer model can be in two different phases,
depending on the strength of the disorder.
For instance, if the effective energies of Eq. \ref{epssite} and \ref{epsbond} are Gaussian
(this corresponds to
 a log-normal distribution for the random fields and for the random bonds
of the quantum models), the Directed Polymer will be in its localized phase
only if the variances of the Gaussians are above some threshold.
Then the quantum model will flow towards an Infinite-Disorder fixed point characterized by
$\omega(d)=\omega_{DP}(D=d-1)$ (see Eq. \ref{omeganumeD} for numerical estimates in dimension $d \leq 8$). Below some disorder threshold, the associated Directed Polymer will be in its delocalized phase, and a Finite-Disorder fixed point for the quantum model is possible.
Note however that in the quantum model, the most frequently used disorder distribution 
is a box distribution for the random fields
\begin{eqnarray}
\pi_{Box}(h) = \frac{1}{\Delta} \theta( 0 \leq h \leq \Delta)
\label{pibox}
\end{eqnarray}
Then the annealed partition function for the effective Directed Polymer model
does not exist ( the inverse moment diverges $\int dh \pi_{Box}(h)/h =+\infty$ as a consequence of the finite
weight at zero-field $\pi_{Box}(h=0)>0 $ ). Then there is no delocalized phase 
 (see section \ref{app_DPfree}) and the Directed Polymer can only be in its localized phase.

\subsection{ Extension to the whole disordered phase   }

\label{extension}

Up to now, we have discussed what happens deep in the disordered phase 
where the typical correlation length $\xi_{typ}$
 introduced in Eq. \ref{xitypomega} is very small.
Let us now discuss what can be extended to the whole disordered phase and what cannot.
As the typical correlation length $\xi_{typ}$ grows, the 'directed constraint' on paths will 
become less and less effective, higher orders of perturbation should be taken into account,
and so on. As a consequence, no quantitative conclusion can be obtained for 
the typical correlation length $\xi_{typ} $,
and in particular, its divergence at the critical point of the quantum model
(where $\delta$ represents some appropriate distance to the quantum transition)
\begin{eqnarray}
\xi_{typ} \propto \delta^{-\nu_{typ}}
\label{xitypnutyp}
\end{eqnarray}
is not determined by the physics of the Directed Polymer.

However, we expect that the fluctuation exponent $\omega$ obtained 
previously deep in the disordered phase
should remain the same in the whole disordered phase of the quantum model.
As a comparison, we describe in Appendix \ref{app_anderson}
 the case of Anderson localization,
where the droplet exponent of the Directed Polymer also appears 
in the localized phase for completely similar reasons, whereas the position of
Anderson transition and the divergence of the localization length cannot
be predicted from the Directed Polymer model.

Let us now discuss the special case of the dimension $d=1$,
where the validity of the perturbative expansion can be analyzed
precisely via a direct comparison with exact result.

\subsection{ Case $d=1$   }

As recalled in the introduction, the quantum model in dimension $d=1$
has been exactly solved by D. Fisher \cite{fisher}  via 
the strong disorder renormalization procedure.
Here to discuss the domain of validity of the above perturbative expansion,
it is however simpler to consider not the magnetization of a center site,
but instead the surface magnetization $m_0^{surf}$ at the boundary $0$, 
when an external magnetization $m_L=m_{ext}$ 
is imposed at the other boundary of the finite chain of length $L$.

\subsubsection{ Surface magnetization at lowest order in perturbation theory in the couplings  }

Within the perturbation theory described above, there exists a single directed path going from
site $0$ to $L$ on the chain, so that the surface magnetization $m_L$ reads
at lowest order in perturbation theory
\begin{eqnarray}
m_0^{surf} \simeq m_{ext} \ \prod_{i=0}^{L-1} \frac{J_{i,i+1}}{h_i} = m_{ext} \exp 
\left[ \displaystyle \sum_{i=0}^{L-1}
\left( \ln J_{i,i+1} - \ln h_i \right) \right]
\label{mperturbd1}
\end{eqnarray}
The logarithm of the magnetization is given by a sum of random variables and 
thus displays the Central Limit scaling 
\begin{eqnarray}
\ln \left( \frac{m_0^{surf}}{  m_{ext}} \right) \opsimeq_{L \to +\infty} 
L \overline{ ( \ln J_{i,i+1} - \ln h_i) } + L^{\frac{1}{2}} u
\label{lnmperturbd1}
\end{eqnarray}
i.e. this perturbative analysis yields that
the fluctuation exponent is simply the random walk exponent
(that corresponds to the degenerate case of a Directed Polymer with $D=0$ transverse
dimensions)
\begin{eqnarray}
\omega(d=1) = \omega_{DP}(D=0)  = \frac{1}{2} 
\label{omegad1}
\end{eqnarray}

\subsubsection{ Comparison with the exact expression for the surface magnetization   }

Let us now compare with the exact expression of the surface magnetization when one imposes $m_L=1$ \cite{msurf},
that can be obtained from a free-fermion representation 
\begin{eqnarray}
m_0^{surf}= \left[ 1+ \sum_{i=0}^{L-1} \prod_{j=0}^i \left( \frac{h_j}{J_{j,j+1} } \right)^2 \right]^{-1/2}
\label{msurfexact}
\end{eqnarray}
At lowest order in the couplings $J_{j,j+1}$, the sum is dominated by the term $i=L-1$ containing the product
of $L$ terms and one recovers the perturbative expression of Eq. \ref{mperturbd1}.
The statistical properties of the exact expression of Eq. \ref{msurfexact}
in both phases and at criticality have been discussed in detail in \cite{msurf,dhar,ckesten}.
The important point is that the fluctuation exponent of Eq. \ref{omegad1}
present in the perturbation theory remains the exact exponent in the whole disordered phase,
as well as the Gaussian distribution of the variable $u$ in Eq. \ref{lnmperturbd1}.
Only the numerical prefactors of the linear term and of the fluctuation terms
are renormalized with respect to the perturbative results.
We note that here the perturbative expression of Eq. \ref{lnmperturbd1} 
actually predicts also
correctly the exact critical point determined by $\overline{ ( \ln J_{i,i+1} - \ln h_i)} =0$
and the typical correlation length exponent $\nu_{typ}=1$,
but we believe that these two last properties are peculiar to the dimension $d=1$, 
and that in higher dimension
$d>1$, the position of the critical point and the exponent $\nu_{typ}$ cannot be obtained
from the Directed Polymer problem (see the discussion in section \ref{extension})

\subsection{ Case of the Cayley tree  }

\label{mcayley}

Let us now consider the case where the quantum model is defined
on a tree of coordination number $(K+1)$. The computation of the center site magnetization $m_0$
in terms of the infinitesimal external magnetization $m_{ext}$ imposed at generation $L$
corresponds at leading order in perturbation theory to the Directed Polymer model
defined on a tree, with the disorder parameters given by Eqs \ref{epssite} and \ref{epsbond} : this corresponds exactly to the 'linearized equations' at zero temperature
obtained after some approximations within the quantum cavity method
 ( Eq 5 of Ref  \cite{ioffe}, Eq 9 of Ref  \cite{feigelman}
or Eq 24 of \cite{dimitrova}).
 We refer to Ref \cite{dimitrova} for the comparison of various levels of approximations.
From the exact solution of the Directed Polymer on a tree \cite{Der_Spo}, 
the droplet exponent vanishes (Eq \ref{omegatree} in Appendix)
\begin{eqnarray}
 \omega_{DP}^{tree}=0
\label{omegatreetext}
\end{eqnarray}
whereas in all low finite dimensions $D=d-1 \leq 7$ that have been tested numerically,
the droplet exponent remains positive $\omega(d)=\omega_{DP}(D)>0$ (Eq. \ref{omeganumeD}).
Since the two cases $\omega>0$ and $\omega=0$ induce very different scaling properties,
we have chosen for clarity to focus on the case of finite $d$ with $\omega(d)>0$
in the following sections,
and to discuss separately the case of the Cayley tree in section \ref{sec_tree}.

\section{ Statistics of the two-point correlation when $\omega(d)>0$ }

\label{sec_corre}

\subsection{ Perturbative expansion deep in the disordered phase of the quantum model }

Deep in the disordered phase of the quantum model, the two-point correlation function 
\begin{eqnarray}
C(\vec r) \equiv < \sigma^z_{\vec 0} \sigma^z_{\vec r}  >
\label{cdef}
\end{eqnarray}
has for leading contribution in perturbation theory (as in Eq. \ref{mperturb})
\begin{eqnarray}
C(\vec r) && =   \sum_{ Path P: {\vec 0} \to {\vec r} } W(P) \nonumber \\
W(P) && = \left( \prod_{(i,j) \in P} \frac{J_{i,j}}{h_i} \right)
\label{cperturb}
\end{eqnarray}
where the sum is over the paths $P$ going from the point $\vec 0$ to the point $\vec r$.
In the disordered phase of the quantum model where the correlation decays exponentially,
we expect again that directed paths dominate the sum, so that the scaling 
of the Directed Polymer model should again appear as Eq. \ref{xitypomega}
\begin{eqnarray}
\ln C(\vec r)  \opsimeq_{ r \to +\infty}  - \frac{r }{\xi_{typ}} +  r^{\omega} u
\label{logcorre}
\end{eqnarray}
This generalizes the known solution in $d=1$ \cite{fisher} where $\omega(d=1)=\omega_{DP}(D=0)=1/2$.
For the quantum model in dimension $d=2$, 
we thus expect $\omega(d=2)=\omega_{DP}(D=1)=1/3$ (Eq. \ref{omegaD1texte}).

Besides the droplet exponent $\omega$ 
already discussed, it is now important to characterize the tail
of the distribution $P(u)$ of the random variable $u$
\begin{eqnarray}
\ln P(u) \opsimeq_{ u \to + \infty} -  u ^{\eta}
\label{etadef}
\end{eqnarray}
From the results recalled in the Appendix, one has the simple relation
 (see the discussion around Eq. \ref{etaomega})
\begin{eqnarray}
\eta(d)= \eta_{DP}(D=d-1)= \frac{1}{1-\omega_{DP}(D=d-1)}=\frac{1}{1-\omega(d)}
\label{etaomegatext}
\end{eqnarray}
that generalizes the exponent $\eta(d=1)=\eta_{DP}(D=0)=2$ of the Gaussian distribution in $d=1$ (Eq. \ref{lnmperturbd1}).

\subsection{ Averaged correlation  }

From the distribution of the logarithm of the two-point correlation of Eq. \ref{logcorre},
one may estimate the contribution of rare events to the averaged correlation :
the correlation $C(r)$ can be of order $O(1)$ if the random variable
$u$ takes the anomalously large value
\begin{eqnarray}
u_{1}  \simeq  + \frac{r^{1-\omega } }{ \xi_{typ}} 
\label{u1}
\end{eqnarray}
and this event occurs with the small probability $P(u_1)$ given by
\begin{eqnarray}
\ln P( u_{1} )  \propto -  u_1^{\eta}  \simeq  -  \frac{r^{(1-\omega)\eta } }{ \xi_{typ}^{\eta} }
\label{pu1}
\end{eqnarray}

Since $(1-\omega)\eta =1$ (Eq \ref{etaomegatext}), this probability of rare events where $C(r) \sim 1$
decays exponentially with the distance $r$, with a characteristic length $ \xi_{typ}^{\eta}$ 
that diverges more rapidly than the typical correlation length $ \xi_{typ}$
whenever the tail exponent satisfies $\eta >1$, i.e. whenever the droplet exponent 
does not vanish $\omega>0$ (Eq \ref{etaomega}).
This means that  the averaged correlation function $C_{av}(r) = \overline{C(r)}
$ will be dominated by these rare events and will decay exponentially 
\begin{eqnarray}
\ln C_{av}(r) \sim  \ln P( u_{1} )  \sim  -  \frac{r }{\xi_{av} } 
\label{cav}
\end{eqnarray}
where the averaged correlation length $\xi_{av}$ 
diverges  as
\begin{eqnarray}
\xi_{av} \propto \xi_{typ}^{\eta} = \xi_{typ}^{ \frac{1}{1-\omega}}
\label{xiav}
\end{eqnarray}
i.e. that there are two distinct correlation length exponents related by
\begin{eqnarray}
\nu_{av} = \nu_{typ} \eta = \frac{ \nu_{typ}}{1-\omega}
\label{nuav}
\end{eqnarray}

In summary, we expect that in finite dimension $d$, where the droplet exponent $\omega_{DP}$ 
of the Directed Polymer does not vanish (see Eq. \ref{omeganumeD}), 
the disordered phase of the quantum model is characterized by
two distinct correlation length exponents related by Eq. \ref{nuav},
that generalize the well-known results $\nu_{typ}=1$ and $\nu_{av}=2$ in $d=1$ \cite{fisher}.

\subsection{ Finite-size scaling of the typical correlation  }

When there exists a single correlation length, this length governs all finite-size properties
near criticality. When the typical and averaged correlation lengths diverge differently,
one expects that it is the averaged correlation length that governs finite-size effects.
It is thus natural to write the following finite-size scaling form for the typical correlation \cite{motrunich}
\begin{eqnarray}
\overline{ \ln C( r)}  \opsimeq_{ r \to +\infty}  - r^{\omega_c} F \left( \frac{r}{\xi_{av} }\right)
\label{fsscorretyp}
\end{eqnarray}
 $\omega_c$ is the exponent governing the decay exactly at criticality
where the scaling function reduces to some positive constant $F(0)>0$.
The matching with the scaling of Eq. \ref{logcorre} in the disordered phase imposes the asymptotic behavior 
\begin{eqnarray}
F(x) \oppropto_{x \to +\infty}  x^{1-\omega_c} 
\label{fscalx}
\end{eqnarray}
leading to 
\begin{eqnarray}
\overline{ \ln C( r)}  \opsimeq_{ r \to +\infty}  -  \frac{r}{\xi_{av}^{1-\omega_c} }
\label{fssmatching}
\end{eqnarray}
i.e. by consistency with Eq. \ref{logcorre}, the typical correlation length should be
$\xi_{typ} \sim \xi_{av}^{1-\omega_c}$. The comparison with Eq. \ref{xiav}
leads to the conclusion
\begin{eqnarray}
\omega_c=\omega
\label{omegac}
\end{eqnarray}
that generalizes the known result $\omega_c=\omega=1/2$ in $d=1$ \cite{fisher}.
In dimension $d=2$, this would give (Eq. \ref{omegaD1texte})
\begin{eqnarray}
\omega_c(d=2)=\omega(d=2)=\frac{1}{3}
\label{omegacd2}
\end{eqnarray}
in agreement with the measure of the critical typical correlation via
Monte-Carlo \cite{pich}.

As a final remark, let us stress 
 that even in dimension $d=1$, the critical behavior of the averaged correlation 
\begin{eqnarray}
\overline{ C_{criti}(r)} \oppropto_{r \to +\infty} r^{- 2 \frac{\beta}{\nu_{av} }}
\label{cavcriti}
\end{eqnarray}
is given by a non-trivial power-law (involving $2 \frac{\beta(d=1)}{\nu_{av}(d=1) }=\beta(d=1)=(3-\sqrt{5})/2$) that cannot be obtained
simply from the Brownian scaling. 
This is because it represents some 'persistence exponent' 
\cite{hastings} for
the strong disorder renormalization, that can only be computed from the exact solution of the renormalization
flow \cite{fisher}. As a consequence, we do not expect any simple exponent for the 
critical averaged correlation in dimension $d>1$ either.

\section{ Statistics of the local susceptibility  when $\omega(d)>0$ }

\label{sec_chi}

\subsection{ Perturbative expansion deep in the disordered phase of the quantum model  }

In the presence of a uniform infinitesimal exterior field $H \to 0$, 
the local magnetizations  $m_i \simeq \chi_{loc}(i) H$ at linear order in $H$
defines the local susceptibilities $\chi_{loc}(i)$.
At first order in the couplings $J_{i,j}$, the local susceptibility $\chi_{loc}(0)$
is related to the local susceptibilities $\chi_{loc}(j)$ 
 of its $K=(2d)$ neighbors on the hypercubic lattice via
\begin{eqnarray}
\chi_{loc}(0) = \frac{1}{h_0} + \sum_{j=1}^{2d} \frac{J_{0,j}}{h_0} \chi_{loc}(j)
\label{chiperturbfirst}
\end{eqnarray}
The difference with Eq \ref{mperturbfirst} is the first term $1/h_0$ that represents
the local susceptibility of the isolated spin $0$ (when $J_{i,j}=0$).
If we iterate Eq \ref{chiperturbfirst}, we obtain
\begin{eqnarray}
\chi_{loc} && =  \sum_{\vec r }   \sum_{ Path P: {\vec 0} \to {\vec r }} W(P) \nonumber \\
W(P) && = \left( \prod_{(i,j) \in P} \frac{J_{i,j}}{h_i} \right)
\label{chilocperturb}
\end{eqnarray}
The difference with Eq \ref{mperturb} is that the path starting at point
 $\vec 0$ now ends at any point $\vec r  $ in the volume $L^d$.
The points $\vec r$ very close to $\vec 0$ give finite contributions, 
whereas the points $\vec r $ far away will have a contribution displaying the scaling 
(see Eqs \ref{mperturb} and  \ref{xitypomega})
\begin{eqnarray}
 \sum_{ Path P: {\vec 0} \to {\vec r }} W(P) \simeq e^{  - \frac{r }{\xi_{typ}} 
+  r^{\omega} u_{\vec r} }
\label{contrirprime}
\end{eqnarray}
where the random variable $ u=u_{\vec r}$ is of order $O(1)$.

\subsection{ Power-law distribution of the local susceptibility  }

A saddle point analysis of Eq. \ref{contrirprime} 
 yields that the leading contribution will come from points $ r $
at a distance
\begin{eqnarray}
r_{saddle} \simeq (\xi_{typ} u)^{\frac{1}{1-\omega}} = \xi_{av} u^{\frac{1}{1-\omega}} 
\label{rprimetyp}
\end{eqnarray}
i.e. of the order of the averaged correlation length $\xi_{av}$ of Eq. \ref{xiav}.
This saddle point gives a contribution of order 
\begin{eqnarray}
\ln \chi_{loc} \sim  + \ \xi_{typ}^{\frac{\omega}{1-\omega}} u^{\frac{1}{1-\omega}} = \xi_{av}^{\omega}
u^{\frac{1}{1-\omega}}
\label{chiloctyp}
\end{eqnarray}
It is thus useful to introduce the positive random variable 
\begin{eqnarray}
 v \equiv \frac{\ln \chi_{loc}}{\xi_{av}^{\omega}} \propto u^{\frac{1}{1-\omega}}=u^{\eta}
\label{defv}
\end{eqnarray}
From the tail of the distribution of the variable $u$ of Eq. \ref{etadef},
one obtains that 
the distribution $Q(v)$ of the variable $v \propto u^{\eta}$ presents the exponential tail (Eq. \ref{etadef} )
\begin{eqnarray}
  Q( v  )  \oppropto_{v \to +\infty}  e^{ - a v} 
\label{qv}
\end{eqnarray}
where $a$ is some constant.
The change of variables from $v$ to $\chi_{loc}$ of Eq. \ref{defv}
 yields that the distribution of the local susceptibility presents the power-law tail 
\begin{eqnarray}
P( \chi_{loc})  \oppropto_{\chi_{loc} \to +\infty}  \frac{ \mu }{  \chi_{loc}^{1+\mu}}
\label{pchilocpower}
\end{eqnarray}
where the exponent $\mu$ varies continuously with the averaged correlation length $\xi_{av}$
\begin{eqnarray}
\mu = \frac{a}{ \xi_{av}^{\omega}}
\label{mudv}
\end{eqnarray}
As the critical point is approached $\xi_{av} \to +\infty$, this exponent $\mu$
goes to zero, and the distribution $P(\chi_{loc})$ becomes broader and broader.
In particular, in the region near the critical point where $0<\mu<1$, the averaged
local susceptibility diverges $\overline{ \chi_{loc}} =+\infty$
(Griffiths phase),
 whereas farther away where $\mu>1$, the averaged value $\overline{ \chi_{loc}} $ remains finite. These conclusions again generalize directly what is known in $d=1$ \cite{fisher}.

 \subsection{ Scaling of the local magnetization as a function of the magnetic field }

In the Griffiths phase $0<\mu<1$ where the averaged local susceptibility diverges
  $\overline{ \chi_{loc}} =+\infty$,
this means that the averaged magnetization 
$\overline{m}$ will grow more rapidly than linearly in the magnetic field $H$.
To see what happens, it is convenient to consider the probability to
have a magnetization $m_0=\chi_{loc} H$ of order $1$ from Eq. \ref{pchilocpower}
\begin{eqnarray}
Prob( m_0= \chi_{loc} H \sim 1) 
= \frac{1}{H} Prob (\chi_{loc} > \frac{1}{H})  \oppropto_{H \to 0} H^{\mu}
\label{H1surz}
\end{eqnarray}
The averaged value  $\overline{ m_0}$ will be thus governed by these rare events
and scales as
\begin{eqnarray}
\overline{ m_0} \sim   \oppropto_{H \to 0} H^{\mu} = e^{ a \frac{ \ln H }{ \xi_{av}^{\omega} }  }
\label{avH1surz}
\end{eqnarray}

In summary, the analysis of this section generalizes the exact results in $d=1$ 
\cite{fisher}
to all finite dimensions $d$ where the droplet exponent remains finite $\omega(d)>0$,
in order to ensure the vanishing of the exponent $\mu$ of Eq. \ref{mudv} near criticality.

\section{ Dynamical scaling  when $\omega(d)>0$ }

\label{sec_dyn}

An essential property of 'conventional' quantum systems is
the dynamical exponent $z$ that describes how the gap $G$ (that defines the
appropriate characteristic inverse time)
of a sample of volume $L^d$ vanishes with $L$ as the power-law
\begin{eqnarray}
G(L^d) \sim L^{-z}
\label{zdef}
\end{eqnarray}
For instance the pure quantum Ising model of Eq. \ref{hpure} is characterized by $z_{pure}=1$
in any dimension $d$, as a consequence of the equivalence with a classical Ising model in dimension
$(d+1)$, where the time plays the role of an extra spatial dimension \cite{sachdev}.
For the disordered model of Eq. \ref{hdes} in dimension $d=1$, the exact solution
\cite{fisher} corresponds however to the following activated scaling at criticality
\begin{eqnarray}
\ln G(L^d) \sim - L^{\psi(d)}
\label{psidef}
\end{eqnarray}
with $\psi(d=1)=1/2$,
instead of the power-law of Eq. \ref{zdef}, 
meaning that the dynamical exponent $z$ is formally infinite at criticality.
In the following, we thus discuss
 the scaling properties of the gap in the disordered phase in dimension $d>1$
where $\omega(d)>0$.

\subsection{ Gap of a sample of volume $L^d$ in the disordered phase }

To relate the study of the gap with the study of the local susceptibility
discussed in the previous section, it is convenient to follow Ref \cite{chiloc}
and to write the local susceptibility 
in terms of the many-body eigenstates $\vert n>$ of energies $E_n$,
where $n=0$ labels the ground state (Eq. 67 of Ref \cite{chiloc})
\begin{eqnarray}
 \chi_{loc}(i) = 2 \sum_{n >0} \frac{\vert < 0 \vert \sigma_i^z \vert n> \vert^2}{E_n-E_0}
\label{chilobmb}
\end{eqnarray}
One expects that $\chi_{loc}$ will behave like the inverse local gap $G_{loc}$,
i.e. more precisely, 
that the local gap $G_{loc}$ in a region of volume (Eq. \ref{rprimetyp}) 
\begin{eqnarray}
V \sim  r_{saddle}^d  \sim \xi_{av}^d
\label{volloctyp}
\end{eqnarray}
behaves as
\begin{eqnarray}
 G_{loc} \sim \frac{1}{\chi_{loc}} 
\label{deltaloctyp}
\end{eqnarray}
The power-law tail of the distribution $P(\chi_{loc})$ of Eq. \ref{pchilocpower}
transforms into the following singularity for the distribution $P(G_{loc})$ 
at the origin
\begin{eqnarray}
P( G_{loc})  \opsimeq_{G \to 0^+}  A \mu G_{loc}^{\mu-1}
\label{pgaplocpower}
\end{eqnarray}
where $A$ is some amplitude. 
Note in particular that in the Griffiths phase $0<\mu<1$, 
this correspond to a diverging singularity at the origin.

As discussed in detail in \cite{FY,extreme}, the gap $G(L^d)$ of the whole system of volume $L^d$
in the disordered phase is given by the minimum value $G^{min}_{loc}$ of the 
$N=L^d/\xi_{av}^d$ independent local gaps $ G_{loc}$ that are present in the system.
Taking into account the power-law singularity of Eq. \ref{pgaplocpower},
one obtains that the gap $G(L^d)=G_{min}$ scales as
\begin{eqnarray}
 \frac{1}{N} = \int_{0}^{G^{min}_{loc}} d G_{loc} P(G_{loc}) \sim  A \left( G^{min}_{loc} \right)^{\mu}
\label{orderwmax}
\end{eqnarray}
i.e. by inversion
\begin{eqnarray}
G(L^d)= G^{min}_{loc}  \propto  \frac{1}{N^{\frac{1}{\mu}}} \sim \left(  \frac{\xi_{av}^d}{L^d} \right)^{\frac{1}{\mu}}
\label{gmin}
\end{eqnarray}
From the dependence in $L$ (Eq \ref{zdef}),  one reads that
the dynamical exponent $z$ is  directly related to the exponent $\mu$ discussed in the previous section
(Eq. \ref{pchilocpower})
\begin{eqnarray}
z= \frac{d}{ \mu} 
\label{zmu}
\end{eqnarray}
This relation involving the factor $d$ \cite{pich,rieger} generalizes the well-known result
$z=1/\mu$ in $d=1$ \cite{fisher}.
In particular the Griffiths phase $0<\mu<1$ corresponds to the domain $z>d$.
The vanishing of the exponent $\mu$ near criticality (Eq \ref{mudv})
yields the following divergence of the dynamical exponent near the transition
\begin{eqnarray}
z= \frac{d}{ \mu} =  \frac{d }{a} \xi_{av}^{\omega}
\label{zmudv}
\end{eqnarray}

\subsection{ Finite-size scaling for the typical gap  }

From the divergence of the dynamical exponent in Eq. \ref{zmudv},
we expect that the critical point
 cannot be conventional but should obey the activated scaling
of Eq. \ref{psidef} with some exponent $\psi$.
In addition, all finite-size-scaling properties
are expected to be governed by the averaged correlation length $\xi_{av}$,
as in Eq. \ref{fsscorretyp} for the typical correlation.
It is thus natural to assume the following finite-size scaling form 
for the typical gap,
in agreement with the exact finite-size scaling form known in $d=1$ \cite{FY}
\begin{eqnarray}
\overline{ \ln G(L^d)}  \opsimeq_{ L \to +\infty}  - L^{\psi} \Phi \left( \frac{L}{\xi_{av} }\right)
\label{fssgaptyp}
\end{eqnarray}
 where the scaling function reduces to some positive constant $\Phi(0)>0$ at criticality.
The matching with the scaling in the disordered phase of Eq. \ref{gmin}
(using the critical behavior of the exponent $\mu$ of Eq. \ref{mudv})
\begin{eqnarray}
 \ln G(L^d) \propto -  \frac{1}{\mu} \ln  \left(  \frac{L^d}{\xi_{av}^d} \right)
\propto -  \xi_{av}^{\omega} \ln  \left(  \frac{L^d}{\xi_{av}^d} \right)
\label{gapdisordered}
\end{eqnarray}
imposes the asymptotic behavior 
\begin{eqnarray}
\Phi(x) \oppropto_{x \to +\infty}  x^{- \psi } \ln (x^d)
\label{phiscalx}
\end{eqnarray}
leading to the conclusion (see also \ref{omegac})
\begin{eqnarray}
\psi=\omega = \omega_c
\label{psiomega}
\end{eqnarray}
that generalizes the known result $\psi=\omega=1/2$ in $d=1$ \cite{fisher}.
In dimension $d=2$, this would give (Eq. \ref{omegaD1texte})
\begin{eqnarray}
\psi(d=2)=\omega(d=2)=\frac{1}{3}
\label{psid2}
\end{eqnarray}
The numerical estimations of $\psi$ tend to be larger, 
around $\psi \simeq 0.42$ \cite{pich,motrunich}
or $\psi \simeq 0.48$ \cite{kovacs2d}, 
whereas the measure of $\omega_c$ of typical critical correlation,
which is supposed to be the same (Eq \ref{omegac}) is closer to $1/3$  \cite{pich}.
Further work is needed to better understand the origin of these differences in $d=2$,
as well as in $d=3,4$ where the measures of the gap exponent 
$\psi$ \cite{kovacs3d,kovacsreview} remain around $0.5$, whereas the droplet exponent $\omega(d)=\omega_{DP}(D=d-1)$ decays with $d$ (see Eq. \ref{omeganumeD}).
A possible explanation for this discrepancy could be that all these
Strong Disorder RG numerical results are based on the approximation of the 'maximum rule'
introduced in the very first paper concerning $d>1$ \cite{motrunich},
that allows to study much bigger sizes.
This 'maximum rule' approximation has been recently questionned \cite{Iyer} for another 
quantum model, namely the superfluid transition for random bosons 
(see the discussions in section III B and Appendix A of \cite{Iyer}).
For our present model, we believe that the use of the approximated 'maximum rules'
instead of the full 'sum rules' could be a problem for the following reasons :

(i) the full 'sum rule' corresponds for the Directed Polymer model to 
the computation of the partition function as a sum over all paths, so that
what shows up is the fluctuation exponent $\omega_{DP}<1/2$ of the free-energy,
that we have discussed in detail in the present paper.

(ii) the approximated 'maximum rule' corresponds for the Directed Polymer model
to the computation of the contribution of a single dominant path, i.e. of the energy of this path.
However at all non-zero temperature $T>0$, the energy and the entropy of the Directed Polymer
are known to display fluctuations of order $L^{1/2}$ in all dimensions $d$,
coming for independent short-scale contributions. There exists a subtle
 'cancellation' between the fluctuating parts of the energy and the entropy at leading
 order, so that the free-energy displays smaller fluctuations of order $L^{\omega_{DP}}<L^{1/2}$
\cite{Fis_Hus_DP} that leads to the phenomenon of 'chaos' in temperature
\cite{Fis_Hus_DP}, as in spin-glasses \cite{Fis_Hus_SG}.

To confirm this interpretation, one needs to study numerically the full 'sum rule',
but unfortunately this seems to be possible only for too small system sizes
to get reliable estimations of the exponents. The remaining possibility
could be to develop other types of approximation, like the cut-off approximation
introduced recently in \cite{Iyer}.

\section{ Case of the Cayley tree  }

\label{sec_tree}

As already mentioned in section \ref{mcayley}, the physics of the 
random field Ising model on the Cayley tree in its disordered phase
has been related to the physics of the Directed Polymer on the Cayley tree in
Refs \cite{ioffe,feigelman,dimitrova} via the quantum cavity method.
In this section, we mention the consequences of this correspondence for various observables
to emphasize the similarities and differences with the case
of finite dimension $d$.
To allow for the possibility of the two phases (delocalized or localized) for the
effective Directed Polymer, we have chosen here not to consider the 'Box'-distribution of Eq. \ref{pibox}
(which is discussed in detail in Refs \cite{ioffe,feigelman,dimitrova}), but to focus instead
on the following log-normal distribution of the random fields
\begin{eqnarray}
\pi_{LN}(h) = \frac{1}{ h \sqrt{2 \pi \sigma^2 }} e^{- \frac{(\ln h - \overline{\ln h})^2}{2 \sigma^2 }}
\label{lognormal}
\end{eqnarray}
of fixed parameters $(\overline{\ln h},\sigma)$, whereas the ferromagnetic couplings $J_{i,j}$ are not random
but take a single value $J$ that will be the control parameter of the quantum transition.

\subsection{Magnetization of a center site   }

On a Cayley tree of coordination number $(K+1)$ with $L$ generations, the local magnetization satisfies the following recurrence at lowest order in perturbation theory 
(analog of Eq. \ref{mperturbfirst})
\begin{eqnarray}
m_L(i) = \frac{J}{h_i} \sum_{j=1}^K   m_{L-1}(j)
\label{recmiDP}
\end{eqnarray}
where $m_{L-1}^{(j)}$ are $K$ independent realizations of the magnetization after $(L-1)$ generations.
This corresponds to the recurrence of the Directed Polymer on a Cayley tree (Eq. \ref{recZDP})
with temperature $T=1$ and with  effective random energies 
(instead of Eqs \ref{epssite} and \ref{epsbond})
\begin{eqnarray}
\epsilon_i= \ln h_i - \ln J
\label{epsi}
\end{eqnarray}
Eq. \ref{lognormal} corresponds to the Gaussian distribution of Eq. \ref{gauss} of averaged value
\begin{eqnarray}
\epsilon_0= \overline{\ln h} - \ln J
\label{epsilonzero}
\end{eqnarray}
The translation of the exact Derrida-Spohn solution \cite{Der_Spo} recalled in section \ref{app_cayley} of the Appendix, yields that
 there exists a critical width $\sigma_c$ 
for the disorder distribution of Eq. \ref{lognormal}
where the critical temperature of Eq. \ref{gausstc} is unity, i.e.
\begin{eqnarray}
\sigma_c =     \sqrt{ 2 \ln K } 
\label{sigmac}
\end{eqnarray}

\subsubsection{ When the Directed Polymer is in its delocalized phase $\sigma<\sigma_c$ }

 The Directed Polymer is in its delocalized phase at $T=1$ if $T=1>T_c$,
 i.e. below the critical disorder $\sigma<\sigma_c$.
Eq. \ref{lnZDPfreeTree} becomes at $T=1$
\begin{eqnarray}
 \ln \left( \frac{m_0}{m_{ext}} \right) \simeq  \ln Z_{DP} (L)
\opsimeq_{L \to +\infty} 
    \left[  \ln  K  +    \frac{ \sigma^2}{2  }  + \ln J  - \overline{\ln h} \right] L
+u
\label{lnZDPfreeTreem}
\end{eqnarray}
where the distribution $ P^{deloc}(u) $ of the random variable $u$ 
 presents the exponential tail  (Eq. \ref{tailufreeTree})
\begin{eqnarray}
 P^{deloc}(u) \oppropto_{u \to +\infty} e^{- \left( \frac{ \sigma_c^2}{\sigma^2} \right) u}  
\label{tailufreeTreemu}
\end{eqnarray}
Equivalently, if one introduces the typical value
 (Eq. \ref{lnZDPfreeTreem} )
\begin{eqnarray}
m_0^{typ} \equiv e^{\overline{ \ln m_0}}
 \oppropto_{L \to +\infty} m_{ext} \ 
 e^{- \left( \overline{\ln h}- \ln  K  -    \frac{ \sigma^2}{2  }  - \ln J \right)L}
\label{motypfreetree}
\end{eqnarray}
one obtains that the distribution
 of the magnetization $m_0$ presents the following power-law tail
\begin{eqnarray}
 {\cal P}^{deloc}(m_0) \oppropto_{m_0 \to +\infty} \frac{ (m_0^{typ})^{\tau_{deloc}} }{ m_0^{1+\tau_{deloc}}} 
\ \ {\rm with } \ \ \tau_{deloc}= \frac{ \sigma_c^2}{\sigma^2} >1 
\label{tailufreeTreem}
\end{eqnarray}
From $\tau_{deloc} >1$, one obtains that the ratio 
 $\overline{m_0}/m_0^{typ}$ remains finite, so that the averaged value has the same 
behavior as the typical value of Eq. \ref{motypfreetree} .

If one assumes that this Directed Polymer approximation 
remains valid in the whole disordered phase
of the quantum model, the behavior of the typical magnetization of Eq. \ref{motypfreetree}
yields that the critical coupling $J_c$ would be
\begin{eqnarray}
\ln J_c  =\overline{\ln h}    - \ln  K  -    \frac{ \sigma^2}{2  }  
\label{jcfree}
\end{eqnarray}
and that the radial correlation length $\xi_{r}$ describing the exponential decay
in the radial direction of the tree would be given for $J<J_c$ by
\begin{eqnarray}
\frac{1}{\xi_{r}} 
= \ln J_c  - \ln J  \oppropto_{J \to J_c^-} (J_c - J)^{\nu_{r} } \ \ {\rm with } \ \ 
\nu_{r}=1
\label{xitypfree}
\end{eqnarray}

\subsubsection{ When the Directed Polymer is in its localized phase $\sigma>\sigma_c$ }

  The Directed Polymer is in its Localized Phase at $T=1$ if $T=1<T_c$, 
i.e. above the critical disorder $\sigma >\sigma_c$.
Eq. \ref{lnZDPdisTree} becomes at $T=1$
\begin{eqnarray}
 \ln \left( \frac{m_0}{m_{ext}} \right) \simeq  \ln Z_{DP} (L)
\opsimeq_{L \to +\infty}   
\left[  \sigma \sigma_c + \ln J  - \overline{\ln h} \right] L
- \frac{3 \sigma }{2 \sigma_c } \ln L
+u
\label{lnZDPdisTreem}
\end{eqnarray}
where the distribution $ P^{loc}(u)$ of the random variable $u$  presents the exponential tail  (Eq \ref{tailudisTree})
\begin{eqnarray}
 P^{loc}(u) \oppropto_{u \to +\infty} u e^{- \frac{\sigma_c}{\sigma} u}  
\label{tailudisTreeu}
\end{eqnarray}
Equivalently, if one introduces the typical value
(Eq. \ref{lnZDPdisTreem})
\begin{eqnarray}
m_0^{typ} \equiv e^{\overline{ \ln m_0}} 
\oppropto_{L \to +\infty} m_{ext} \ L^{- \frac{3 \sigma }{2 \sigma_c } } \ e^{- \left( \overline{\ln h} - \sigma \sigma_c - \ln J  \right)L} 
\label{motypdistree}
\end{eqnarray}
one obtains that
the distribution
 of the magnetization $m_0$ presents the following power-law tail
(with a logarithmic correction)
\begin{eqnarray}
 {\cal P}^{loc}(m_0) \oppropto_{m_0 \to +\infty}
 \frac{ (m_0^{typ})^{\tau_{loc}} }{ m_0^{1+\tau_{loc}}} \ \ln \left( \frac{m_0}{m_0^{typ}} \right)
\ \ {\rm with } \ \ \tau_{loc}= \frac{ \sigma_c}{\sigma} < 1 
\label{tailulocTreem}
\end{eqnarray}
From $\tau_{loc} <1$, one obtains that the ratio $\overline{m_0}/m_0^{typ}$ diverges.

If one assumes that this Directed
 Polymer approximation remains valid in the whole disordered phase
of the quantum model, the typical magnetization of Eq. \ref{motypdistree}
yields the critical coupling $J_c$ would be
\begin{eqnarray}
\ln J_c  =\overline{\ln h}   - \sigma \sigma_c  
\label{jcloc}
\end{eqnarray}
and that the radial correlation length $\xi_{r}$  for $J<J_c$ would be 
\begin{eqnarray}
\frac{1}{\xi_{r}} 
 = \ln J_c  - \ln J 
\oppropto_{J \to J_c^-} (J_c - J)^{\nu_{r} } \ \ {\rm with } \ \ 
\nu_{r}=1
\label{xityploc}
\end{eqnarray}

\subsubsection{ Possible finite-size scaling }

Critical points of disordered models on the Cayley 
tree usually involve two correlation length exponents,
as shown by the examples of 
 the localization/delocalization transition of the Directed Polymer
\cite{Cook_FSS,us_critiDP}, 
the random wetting transition \cite{us_wettingCayley}
and of Anderson transition \cite{us_andersonCayley}.
From the analogy with the Anderson transition, where the localized phase
can be studied by a traveling-wave approach, and where the transition corresponds
to a traveling-non traveling phase transition, 
one might expect the following finite-size scaling form
for the typical center-site magnetization
\begin{eqnarray}
 \overline{ \ln \left( \frac{m_0}{m_{ext}} \right) }
\opsimeq_{L \to +\infty}   - L^{\rho} \Phi \left( L^{1/\nu_{FS}} (J_c-J)  \right)
\label{fssm0}
\end{eqnarray}
with some finite-size scaling exponent $\nu_{FS}$.
This type of finite-size scaling for a
traveling/non-traveling phase transition has been found in Ref \cite{simon}
for a soluble case with explicit critical exponents, and for the Anderson transition
with numerical estimates of the exponents \cite{us_andersonCayley}.
The matching with the disordered phase
 involving the radial correlation length $\xi_r \propto (J_c-J)^{- \nu_r}$
discussed previously
\begin{eqnarray}
 \overline{ \ln \left( \frac{m_0}{m_{ext}} \right) } 
\opsimeq_{J<J_c}   - \frac{L}{\xi_r} \simeq - L  (J_c-J)^{\nu_r}
\label{fssm0matchingdis}
\end{eqnarray}
leads to the relation
\begin{eqnarray}
\rho+\frac{\nu_r}{\nu_{FS}} =1 
\label{fssm0relationdis}
\end{eqnarray}
In the ordered phase $J>J_c$, 
one might expect in analogy with \cite{simon,us_andersonCayley}
the essential singularity
\begin{eqnarray}
 \overline{ \ln \left( \frac{m_0}{m_{ext}} \right) }
\opsimeq_{L \to +\infty}   -  (J-J_c)^{-\kappa}  
\label{fssm0matchingordered}
\end{eqnarray}
The matching with Eq. \ref{fssm0} gives the relation
\begin{eqnarray}
\kappa= \rho \nu_{FS}
\label{fssm0relationorder}
\end{eqnarray}

Within our present approach, we can use the associated Directed Polymer
model only in the quantum disordered phase, but not at criticality.
As a consequence, we cannot justify the behavior at criticality or the essential
singularity of Eq. \ref{fssm0matchingordered}.
Using some approximations within the quantum cavity method
(which may however break down close to criticality),
Refs \cite{ioffe,feigelman,dimitrova} have found the essential singularity
of Eq. \ref{fssm0matchingordered} with $\kappa=1$. Taking into account $\nu_r=1$
(Eqs \ref{xityploc}), this would correspond to the simple values
$\nu_{FS}=2$ and $\rho=1/2$.

As a final remark, we note that for the Anderson transition,
the exponent $\kappa=1/2$ of the essential singularity has been exactly computed
 \cite{MFdc=infty}

\subsection{  Statistics of the local susceptibility in the disordered phase }

On the Cayley tree, the recurrence at lowest order for the local susceptibilities
read (Eq. \ref{chiperturbfirst})
\begin{eqnarray}
\chi_{loc}^{(L)}(i) = \frac{1}{h_i} + \frac{ J }{ h_i}   \sum_{j=1}^K  \chi_{loc}^{(L-1)}(j)
\label{recchiDP}
\end{eqnarray}
where $\chi_{loc}^{(L-1)}(j)$ are $K$ independent realizations of the local susceptibilities
 after $(L-1)$ generations.
Without the inhomogeneous term $1/h_i$, 
one would recover the recurrence of Eq. \ref{recmiDP}
for the magnetizations $m_i$ that decay exponentially towards zero in the disordered phase.
As a consequence, one expects that the recurrence of Eq. \ref{recchiDP} converges towards
a finite random variable $\chi_{loc}= \chi_{loc}^{(L \to +\infty)}$
 that should be stable upon iteration.
Upon iteration, Eq. \ref{chilocperturbtree} yields that the local susceptibility
can be seen of a sum over partition functions of the Directed Polymer on the tree
of arbitrary length $r$ (analog of Eq \ref{chilocperturb})
\begin{eqnarray}
\chi_{loc} \sim  \sum_{r }  Z_{DP} (r)
\label{chilocperturbtree}
\end{eqnarray}

Let us discuss the behavior of the averaged value $\overline{\chi_{loc}}$,
the behavior of the typical value $\chi_{loc}^{typ}$,
and then the tail of the probability distribution $P(\chi_{loc})$.

\subsubsection{  Behavior of the averaged value  }

Averaging Eq. \ref{recchiDP} over the disorder
leads to the following closed expression for the averaged value
$\overline{\chi_{loc}}$, that we may explicitly compute using the 
disorder distribution of Eq. \ref{lognormal}, and the value of $J_c$ of Eq. \ref{jcfree}
\begin{eqnarray}
\overline{ \chi_{loc}}  = \frac{ \int dh \frac{\pi_{LN}(h)}{h} }
 {1- K J \int dh \frac{\pi_{LN}(h)}{h} }  = 
\frac{ e^{\frac{\sigma^2}{2}- \overline{\ln h}} }{1- K J  e^{\frac{\sigma^2}{2}- \overline{\ln h}}}
=  \frac{ 1 }{ K (J_c -  J )}
\label{chiavDP}
\end{eqnarray}
i.e. it diverges with the 'mean-field' exponent $\gamma=1$.

\subsubsection{  Typical value for $\sigma<\sigma_c$  }

For $\sigma<\sigma_c$ where Eq. \ref{lnZDPfreeTreem} holds
\begin{eqnarray}
Z_{DP} (r) \simeq e^{ - \frac{r}{\xi_r} +u }
\label{lnZDPfreeTreemforchi}
\end{eqnarray}
we expect that the typical local correlation scales as the radial correlation length
$\xi_r$
\begin{eqnarray}
\chi_{loc}^{typ} \sim  \int dr   Z_{DP} (r) \sim  \int dr e^{ - \frac{r}{\xi_r} } \propto \xi_r \propto (J_c-J)^{-\nu_{typ}}
\label{chiloctypfree}
\end{eqnarray}
Taking into account $\nu_{typ}=1$, one obtains that the typical value diverges with
the same mean-field exponent $\gamma_{typ}=1$ as the averaged value of Eq. \ref{chiavDP}.

\subsubsection{  Typical value for $\sigma>\sigma_c$  }

For $\sigma>\sigma_c$ where Eq. \ref{lnZDPdisTreem} holds
\begin{eqnarray}
Z_{DP} (r) \simeq r^{- \frac{3 \sigma }{2 \sigma_c }} \ \ e^{ - \frac{r}{\xi_r} +u }
\label{lnZDPdisTreemforchi}
\end{eqnarray}
one obtains the following contribution
\begin{eqnarray}
\chi_{loc}^{typ} \sim \int dr   Z_{DP} (r) \sim  \int dr  r^{- \frac{3 \sigma }{2 \sigma_c }} e^{ - \frac{r}{\xi_r} } 
\propto \xi_r^{1 - \frac{3 \sigma }{2 \sigma_c }} 
\label{chiloctypfrozen}
\end{eqnarray}
that does not diverge with $\xi_r$ since $\sigma>\sigma_c$.
This means that the typical value $\chi_{loc}^{typ} $ remains finite
as the transition is approached $\xi_r \to +\infty$, as a consequence of 
the value of the coefficient $\frac{3 \sigma }{2 \sigma_c }>1 $ of the logarithmic correction of the Directed Polymer model (Eq. \ref{lnZDPdisTree}) !

\subsubsection{  Exponent $\mu$ of the power-law distribution  }

From the recurrence of Eq. \ref{recchiDP}, 
one expect that the stable distribution $P(\chi_{loc})$ will display a power-law tail
\begin{eqnarray}
P(\chi_{loc}) \opsimeq_{\chi_{loc} \to +\infty} \frac{A (\chi_{loc}^{typ})^{\mu}}{\chi_{loc}^{1+\mu}}
\label{defmutree}
\end{eqnarray}
where $A$ is some amplitude.
Note that for the special case $K=1$ where the tree degenerates into a single line,
the recurrence of Eq. \ref{recchiDP} defines a so-called Kesten variable
that has been much studied in various physical contexts
 \cite{Kesten,Der_Pom,Bou,Der_Hil,Cal} with the conclusion that
$\mu$ is determined by the condition (written for our present case) 
\begin{eqnarray}
 { \rm Case  } \ K=1 \ \ : \ \ \overline{\left( \frac{J}{h_i} \right)^{\mu}}=1
\label{mukesten}
\end{eqnarray}
For the tree with $K>1$, it is straightforward to adapt the argument as follows.
The distribution $P(\chi_{loc})$ has to be stable upon the iteration
 of Eq. \ref{recchiDP} when the random field $h_i$ is distributed with $\pi_{LN}(h) $
\begin{eqnarray}
P(\chi_{loc}) = \int dh \pi_{LN}(h) \int d\chi_{loc}^{(1)} P(\chi_{loc}^{(1)})
\int d\chi_{loc}^{(2)} P(\chi_{loc}^{(2)}) .. \int d\chi_{loc}^{(K)} P(\chi_{loc}^{(K)})
\delta \left( \chi_{loc}- \frac{1}{h} - \frac{ J }{ h}   \sum_{j=1}^K  \chi_{loc}^{(j)}\right)
\label{eqmsp}
\end{eqnarray}
The stability of the power-law tail of Eq \ref{defmutree}
in the region $\chi_{loc} \to +\infty$ can be analyzed as follows :
a very large $\chi_{loc}$ is obtained only if one of the $K$ values
$\chi_{loc}^{(j)} $ is also large; assuming it is $j=1$, one has then $\chi_{loc}
\simeq \frac{ J }{ h} \chi_{loc}^{(j)}$ in the delta function, so that 
Eq \ref{eqmsp} becomes in the tail region
\begin{eqnarray}
\frac{A (\chi_{loc}^{typ})^{\mu} }{\chi_{loc}^{1+\mu}} &&  \simeq K
 \int dh \pi_{LN}(h) \int d\chi_{loc}^{(1)} \frac{A (\chi_{loc}^{typ})^{\mu}}{(\chi_{loc}^{(1)})^{1+\mu}}
\delta \left( \chi_{loc}- \frac{ J }{ h}   \chi_{loc}^{(1)}\right) \nonumber \\
&& \simeq K J^{\mu} 
 \int dh \pi_{LN}(h) h^{-\mu}  \frac{A(\chi_{loc}^{typ})^{\mu}}{(\chi_{loc})^{1+\mu}}
\label{tailstability}
\end{eqnarray}
yielding that the tail exponent $\mu$ is stable only if it satisfies
\begin{eqnarray}
 K \overline{\left( \frac{J}{h_i} \right)^{\mu}} =1
\label{eqstabmu}
\end{eqnarray}
that generalizes Eq. \ref{mukesten} to arbitrary $K$.

With the disorder distribution of Eq. \ref{lognormal}, the stability condition of Eq. \ref{eqstabmu}
reads 
\begin{eqnarray}
1= K e^{ \frac{\mu^2 \sigma^2}{2} - \mu ( \overline{\ln h}- \ln J ) }
\label{eqstabmuLN}
\end{eqnarray}
Taking the appropriate solution of this second-order equation in $\mu$, one finally 
obtains
\begin{eqnarray}
\mu = \frac{1}{\sigma^2} \left[ ( \overline{\ln h}- \ln J )
+ \sqrt{( \overline{\ln h}- \ln J )^2 - 2 \sigma^2 \ln K } \right]
\label{solmu}
\end{eqnarray}
that generalizes the known result $\mu =\frac{2}{\sigma^2} 
( \overline{\ln h}- \ln J ) $ for the one-dimensional case $K=1$ of Kesten variables.
To see more clearly how the exponent $\mu$ varies within the 
quantum disordered phase as a function of the radial correlation
length $\xi_r$, let us now distinguish the two cases :

(i) For $\sigma<\sigma_c$, using Eqs \ref{sigmac} and \ref{xitypfree},
one obtains
\begin{eqnarray}
\mu_{deloc} = \frac{1}{\sigma^2} 
\left[ \left( \frac{ \sigma^2+\sigma_c^2}{2  } + \frac{1}{\xi_{r}} \right)
+ \sqrt{\left(    \frac{ \sigma^2+\sigma_c^2}{2  } + \frac{1}{\xi_{r}} \right)^2 
-  \sigma^2 \sigma_c^2 } \right]
\label{solmufree}
\end{eqnarray}
The exponent $\mu_{deloc}$ decreases as $\xi_r$ grows
and the limit value near criticality $\xi_r \to +\infty$ 
\begin{eqnarray}
\mu_{deloc} \opsimeq_{\xi_r \to +\infty}  \frac{1}{\sigma^2} 
\left[  \frac{ \sigma^2+\sigma_c^2}{2  } 
+ \sqrt{ \left(  \frac{ \sigma_c^2 - \sigma^2}{2  } \right)^2 } \right] 
= \frac{\sigma_c^2}{\sigma^2}
\label{solmufreecriti}
\end{eqnarray}
coincides with the exponent $\tau_{deloc}>1$ of Eq. \ref{tailufreeTreem}.

(ii) For $\sigma>\sigma_c$ using Eqs \ref{sigmac} and \ref{xityploc}
\begin{eqnarray}
\mu_{loc} = \frac{1}{\sigma^2} \left[ \left( \sigma \sigma_c + \frac{1}{\xi_{r}} \right)
+ \sqrt{\left( \sigma \sigma_c + \frac{1}{\xi_{r}} \right)^2 -  \sigma^2 \sigma_c^2 } \right]
\label{solmufrozen}
\end{eqnarray}
The exponent $\mu_{loc}$ decreases as $\xi_r$ grows
and the limit value near criticality $\xi_r \to +\infty$ 
\begin{eqnarray}
\mu_{loc} \opsimeq_{\xi_r \to +\infty} 
 \frac{1}{\sigma^2} \left[  \sigma \sigma_c  \right] = \frac{\sigma_c}{\sigma} 
\label{solmufrozencriti}
\end{eqnarray}
coincides with the exponent $\tau_{loc}<1$ of Eq. \ref{tailulocTreem}.

In summary, for the log-normal distribution of the random fields (Eq. \ref{lognormal}),
the statistics of the local susceptibility for $\sigma>\sigma_c$
is very similar to the results found
in \cite{ioffe,feigelman,dimitrova} for the case of the Box Distribution
 of Eq. \ref{pibox} :
the exponent $\mu_{loc}$ converges towards a finite value $\frac{\sigma_c}{\sigma}<1$
as the critical point is approached $\xi_r \to +\infty$, instead of converging
towards zero when $\omega(d)>0$ in finite dimension $d$ (Eq. \ref{mudv}).

\subsection{  Dynamical scaling  in the disordered phase }

As in section \ref{sec_dyn}, the power-law distribution of the local susceptibility
(Eq. \ref{defmutree}) leads to a power-law singularity for the distribution
$P( G_{loc})$ of the local gap $G_{loc}$ 
(see the discussion leading to Eq. \ref{deltaloctyp})
\begin{eqnarray}
P( G_{loc})  \opsimeq_{G \to 0^+}  A (G_{loc}^{typ})^{-\mu} G_{loc}^{\mu-1}
\label{pgaplocpowertree}
\end{eqnarray}
In a Cayley tree with $L$ generations of volume $K^L$,
the number $N$ of independent local gaps $ G_{loc}$ will also grow exponentially in $L$.
As a consequence, the gap $G(K^L)$ of the whole system 
in the quantum disordered phase that is given by the minimum value $G^{min}_{loc}$ of the 
$N$ independent local gaps $ G_{loc}$ is expected to scale as
\begin{eqnarray}
\ln G(K^L)= \ln G^{min}_{loc}  \propto  \ln \left( \frac{1}{N^{\frac{1}{\mu}}} \right)
\propto - \frac{1}{\mu} \ln N 
\propto   - L 
\label{gmintree}
\end{eqnarray}
instead of Eq. \ref{gmin} in finite dimension $d$.
So the gap $G(K^L) $ decays only exponentially with respect to the radial size $L$,
and this corresponds formally to an activated exponent $\psi=1$ 
and to an infinite dynamical exponent $z=+\infty$ already within the quantum disordered
phase, in agreement with the limit $d \to +\infty$ of Eq. \ref{zmu} at fixed $\mu$.

\section{ Conclusion  }

\label{sec_conclusion}

For the random transverse field Ising model in finite dimensions $d>1$, we have analyzed the lowest-order contributions in perturbation theory in $(J_{i,j}/h_i)$ to obtain some information on the statistics of various observables in the disordered phase. We have established a link with the statistics of the partition function of the Directed Polymer with $D=(d-1)$ transverse directions. We have analyzed the consequences in terms of the droplet exponent $\omega=\omega_{DP}(D=d-1)$ of the associated Directed Polymer with the following conclusions :

(i) Whenever the Directed Polymer is in its localized phase with a positive droplet
exponent $\omega>0$, the quantum model is governed by an Infinite-Disorder fixed point.
In particular, there are two distinct correlation length exponents related by $\nu_{typ}=(1-\omega)\nu_{av}$. The distribution of the local susceptibility $\chi_{loc}$ presents the power-law tail $P(\chi_{loc}) \sim 1/\chi_{loc}^{1+\mu}$  where $\mu$ vanishes as $\xi_{av}^{-\omega} $, so that the averaged local susceptibility diverges in a finite neighborhood $0<\mu<1$ before criticality (Griffiths phase).
The dynamical exponent $z$ diverges near criticality as $z=d/\mu \sim \xi_{av}^{\omega}$. 

(ii) In dimensions $d \leq 3$, the associated Directed Polymer
 is always in its localized phase (the delocalized phase does not exist), so that any infinitesimal disorder flows towards this Infinite-Disorder fixed point with $\omega(d)>0$ (for instance $\omega(d=2)=1/3$ and $\omega(d=3) \sim 0.24$) 

(iii) In finite dimensions $d > 3$, the associated Directed Polymer can be in two phases
depending on the disorder strength :

(iiia) For strong enough disorder, the flow is towards an Infinite-Disorder fixed point 
provided the droplet exponent of the localized phase of the associated Directed Polymer
remains positive $\omega=\omega_{DP}(D=d-1)>0$ (this seems the case for at least $d \leq 8$ from the numerical results quoted in Eq. \ref{omeganumeD} and it could be for all $d<+\infty$ if the upper critical dimension is infinity for the Directed Polymer model).

(iiib) For small enough disorder, the associated Directed Polymer will be in its delocalized phase with $\omega_{deloc}=0$, so that the quantum model could then flow towards a more conventional Finite-Disorder fixed point.

\vskip 1cm

{ \it Note added in proof : }

In this paper, we have presented the simplest scenario
where the random variable u of Eq.  7 or 20 remains of order O(1)
as in the exact solution in dimension d=1 and as found in dimension d=2 via a
non-linear transfert approach [C. Monthus ans T. Garel, J. Stat. Mech. (2012)
P01008].
However, a more complicated scenario where the random variable u of Eq 7 or 20
contains a diverging amplitude as the critical point is approached, is also
possible, as was found recentlty on a hierarchical fractal lattice [C. Monthus
ans T. Garel, arxiv:1201.6136].
We refer to this preprint for the description of the consequences on various
scalings and on relations between critical exponents.

\appendix

\section{ Reminder on the Directed Polymer in a random medium }

\label{app_DP}

 The model of the Directed Polymer in a $(1+D)$ random medium 
(where $D$ is the dimension of transverse spatial directions,
and where the $1$ refers to the 'directed' direction)
 is defined by the following classical partition function at inverse temperature $\beta=1/T$
 \begin{equation}
Z^{DP}_{L}(\beta) = \displaystyle \sum_{RW} 
\exp \left( - \beta \displaystyle \sum_{1 \leq \alpha \leq L} 
\epsilon (\alpha, \vec r(\alpha))  \right) 
\label{directed}
  \end{equation}
The sum is 
over $D-$dimensional random walks $\vec r(\alpha)$, where the effective 'time' $\alpha$ represents the directed direction.
The independent random energies $\epsilon (\alpha, \vec r)$ can be put either on the sites
or on the bounds.
This model has attracted a lot of attention because it is directly related
to non-equilibrium properties of growth models in the KPZ universality class 
\cite{Hal_Zha}.
Within the field of disordered systems, it is also very interesting on its own
because it represents a `baby-spin-glass' model
\cite{Hal_Zha,Der_Spo,Der,Mez,Fis_Hus_DP}. 

The possible phases of the Directed Polymer are :

(i) a localized phase at low temperature $T<T_c$, where the 
order parameter is an `overlap' \cite{Der_Spo,Mez}.
In finite dimensions, a scaling droplet theory was proposed \cite{Fis_Hus_DP},
in direct correspondence with the droplet theory of spin-glasses \cite{Fis_Hus_SG},
whereas in the mean-field version of the model on the Cayley,
the localized phase is very similar to the frozen phase occurring
in the Random Energy Model \cite{Der_Spo}.

(ii) a delocalized phase at high temperature $T>T_c$

In the following, we recall the properties of these phases that are needed in the text,
 as well as the phase diagram as a function of the dimension $D$

\subsection{ Properties of the  localized phase }

\label{app_DPloc}

\subsubsection{ Droplet exponent $\omega_{DP}(D)$  }

The droplet theory for Directed Polymers \cite{Fis_Hus_DP}
is similar to the droplet theory of spin-glasses \cite{Fis_Hus_SG}.
It is a scaling theory that can be summarized as follows.
At very low temperature $ T \to 0$, all observables are governed by
the statistics of low energy excitations above the ground state.
An excitation of large length $l$ costs a random energy
\begin{eqnarray}
 \Delta E(l) \sim l^{\omega_{DP}} 
\label{ground}
\end{eqnarray}
The droplet exponent $\omega_{DP}$ 
is exactly known in dimension $D=1$  \cite{Hus_Hen_Fis,Kar,Joh,Pra_Spo}
\begin{eqnarray}
 \omega_{DP}(D=1)=\frac{1}{3}
\label{omegaD1}
\end{eqnarray}
and for the mean-field version on the Cayley tree
  \cite{Der_Spo}
\begin{eqnarray}
 \omega_{DP}^{tree}=0
\label{omegatree}
\end{eqnarray}
In finite dimensions $D=2,3,4,5,...$, the exponent $\omega_{DP}(D)$ has
been numerically measured, in particular  \cite{Tan_For_Wol,Ala_etal,perlsman,KimetAla,Mar_etal,DPtails,schwartz}
\begin{eqnarray}
 \omega_{DP}(D=2) && \simeq 0.244  \nonumber \\
 \omega_{DP}(D=3) && \simeq 0.186  \nonumber \\
\omega_{DP}(D=4) && \simeq 0.153  \nonumber \\
\omega_{DP}(D=5) && \simeq 0.130  \nonumber \\
\omega_{DP}(D=6) && \simeq 0.114  \nonumber \\
\omega_{DP}(D=7) && \simeq 0.100
\label{omeganumeD}
\end{eqnarray}
The authors of Ref. \cite{Mar_etal} have moreover argued
that the droplet exponent remains positive $\omega_{DP}(D)>0$ in all finite dimension $D$.

\subsubsection{ Distribution of the partition function  : tail exponent $\eta$ }

One expects that the whole low temperature phase $0<T<T_c$
is governed by the zero-temperature fixed point characterized by the droplet exponent $\omega_{DP}$.
In particular, the logarithm of the partition function scales for $T<T_c$ as
\begin{eqnarray}
\ln Z_L \opsimeq_{L \to +\infty}  - \beta f_{loc} L +  L^{\omega_{DP}} u
\label{lnZDPloc}
\end{eqnarray}
The first term is extensive and non-random
($f_{loc}$ represents the free-energy per step in this localized phase).
The second term involving the droplet exponent $\omega_{DP}$ is random :
$u$ is a random variable distributed with some law $p(u)$,
that presents the following asymptotic behavior 
\begin{eqnarray}
\ln p(u) \oppropto_{u \to +\infty}  -  u^{\eta}
\label{puDP}
\end{eqnarray}
where the tail exponent $\eta(D)$ is directly related to the droplet exponent $\omega_{DP}(D)$ via
(see \cite{DPtails,diamondtails,matching} and references therein)
\begin{eqnarray}
\eta_{DP} = \frac{1}{1- \omega_{DP}}
\label{etaomega}
\end{eqnarray}

In dimension $D=1$, the distribution $p(u)$ is exactly known to be the Tracy-Widom distribution 
\cite{Joh,Pra_Spo,Dot,Cala,sasa,corwin} with the tail exponent
\begin{eqnarray}
\eta_{DP}(D=1) = \frac{3}{2}
\label{etaD1}
\end{eqnarray}
On the Cayley tree, the distribution $p(u)$ is also exactly known  \cite{Der_Spo}
with the tail exponent
\begin{eqnarray}
\eta_{DP}^{tree} = 1
\label{etatree}
\end{eqnarray}

\subsection{ Properties of the delocalized phase }

\label{app_DPfree}

In the delocalized phase $T>T_c$ , the logarithm of the partition function scales as
(compare with Eq. \ref{lnZDPloc})
\begin{eqnarray}
\ln Z_L \opsimeq_{L \to +\infty}  - \beta f_{deloc} L +  u
\label{lnZDPdeloc}
\end{eqnarray}
In the extensive non-random term, 
 the free-energy per step $f_{deloc}$ is given by the annealed value $f_{ann}$
that can be obtained from the averaged partition function $\overline {Z_L}$
\begin{eqnarray}
  - \beta f_{deloc} = - \beta f_{ann} = \oplim_{L \to +\infty} \left( \frac{\ln (\overline {Z_L} ) }{L}\right) 
\label{fann}
\end{eqnarray}
The second term in Eq. \ref{lnZDPdeloc} is a random variable 
$u$ of order $L^0$.

\subsection{ Possible phases as a function of the number $D$ of transverse directions }

\label{app_phasediagram}

The phase diagram as a function of space dimension $D$ is the
following \cite{Hal_Zha} :

(i)  In dimension $D \leq 2$, there is no delocalized phase 
i.e. the critical temperature is infinite 
\begin{eqnarray}
 T_c(D \leq 2) = +\infty 
\label{tcinfinite}
\end{eqnarray}
and any initial disorder drives the polymer into the localized phase.

(ii) In dimension $D>2$, there can be a phase transition between
the low temperature localized phase
and a delocalized phase at high temperature  \cite{Imb_Spe,Coo_Der}.
 This phase transition  has been studied exactly on a Cayley tree \cite{Der_Spo}. 
In finite dimensions, bounds on the critical temperature
$T_c$ have been derived \cite{Coo_Der,Der_Gol,Der_Eva} :
$T_0 \le T_c \le T_2$.
The upper bound $T_2$ corresponds to the temperature above which the ratio
$\overline{Z_L^2}/(\overline{Z_L})^2$ remains finite as $L \to
\infty$. The lower bound $T_0$ corresponds to the temperature below which
the annealed entropy becomes negative. 
For instance, when the random energies in Eq. \ref{directed} are Gaussian variables
\begin{eqnarray}
\rho(\epsilon) = \frac{1}{\sqrt{2 \pi \sigma^2 }} e^{- \frac{(\epsilon-\epsilon_0)^2}{2 \sigma^2 }}
\label{gauss}
\end{eqnarray}
the temperature $T_2$ can be exactly computed and is finite for $D \geq 3$ \cite{Der_Gol},
so that one is sure that the delocalized phase exists at least in the region $T>T_2$.

\subsection{ Reminder on the exact solution on the Cayley tree \cite{Der_Spo} }

\label{app_cayley}

The Directed Polymer model on the Cayley tree of coordination number $(K+1)$
can be solved for any disorder distribution via the Derrida-Spohn traveling-wave approach
\cite{Der_Spo} for the recurrence
\begin{eqnarray}
Z_L(i) = e^{- \beta \epsilon_i}  \sum_{j=1}^K  Z_{L-1}(j)
\label{recZDP}
\end{eqnarray}
where $Z_{L-1}^{(j)}$ are $K$ independent realizations of the partition function after $(L-1)$ generations.
Here we recall the solution for the Gaussian distribution of Eq. \ref{gauss}
that leads to very simple explicit results \cite{Der_Spo} :
the annealed partition function reads
\begin{eqnarray}
\overline{ Z_L } = \left( K \overline{e^{- \beta \epsilon }} \right)^L = 
\left( K e^{ \beta \left( \frac{\beta \sigma^2}{2}-\epsilon_0\right) } \right)^L
\label{gausszann}
\end{eqnarray}
so that the annealed free-energy per step at temperature $T=1/\beta$ reads
\begin{eqnarray}
f_{ann}(T) \equiv \oplim_{L \to +\infty} \left( - \frac{ \ln \left( \overline{ Z_L } \right)  }{\beta L} \right)
 = \epsilon_0 - T \ln  K -    \frac{ \sigma^2}{2 T } 
\label{gausszanfannT}
\end{eqnarray}
with the corresponding annealed entropy
\begin{eqnarray}
s_{ann}(T) \equiv - \partial_{T} f_{ann}(T)  =   \ln  K -   \frac{ \sigma^2}{2 T^2 } 
\label{gausszansannT}
\end{eqnarray}
The transition temperature $T_c$ corresponds
 to the temperature where the annealed entropy vanishes $s_{ann}(T_c)=0$ so
\begin{eqnarray}
T_c =      \frac{ \sigma}{ \sqrt{ 2 \ln K } } 
\label{gausstc}
\end{eqnarray}
and the two phases have the following properties

 In the Delocalized Phase $T>T_c$, the free-energy coincides with the annealed
 free-energy of Eq. \ref{gausszanfannT}, 
and the logarithm of the partition function has for statistics 
\begin{eqnarray}
 \ln Z_L \opsimeq_{L \to +\infty} 
     \left[ \ln  K  +    \frac{ \sigma^2}{2 T^2 }  - \frac{ \epsilon_0}{T} \right] L +u
\label{lnZDPfreeTree}
\end{eqnarray}
where the random variable $u$
 of order $O(1)$ has a distribution $P^{deloc}(u) $
 that presents the exponential tail  \cite{Der_Spo}
\begin{eqnarray}
 P^{deloc}(u) \oppropto_{u \to +\infty} e^{- \left( \frac{ T^2}{T_c^2} \right) u}  
\label{tailufreeTree}
\end{eqnarray}

  In the Localized Phase $T<T_c$, the free-energy per step remains frozen at the value $f_{ann}(T_c)$,
 and the logarithm of the partition function has for statistics 
\begin{eqnarray}
\ln Z_L \opsimeq_{L \to +\infty} 
=   \left[ \frac{\sigma\sqrt{ 2 \ln K }}{T }- \frac{\epsilon_0}{T}  \right] L
- \frac{3 T_c}{2 T} \ln L
+u
\label{lnZDPdisTree}
\end{eqnarray}
where the first correction to the extensive term is a non-random logarithmic term
\cite{Der_Spo,Cook_FSS}, and
where the variable $u$ 
 of order $O(1)$ has distribution $ P^{loc}(u)$
that presents the exponential tail  \cite{Der_Spo}
\begin{eqnarray}
 P^{loc}(u) \oppropto_{u \to +\infty} u e^{- \frac{T}{T_c} u}  
\label{tailudisTree}
\end{eqnarray}

\section{ Analogy with the localized phase of Anderson localization  }

\label{app_anderson}

 The Anderson tight-binding model for a single quantum particle on a hypercubic lattice in dimension $d$
is defined in terms of the Hamiltonian
\begin{eqnarray}
H = \sum_i \epsilon_i \vert i > < i \vert 
 +  \sum_{i,j}  V_{i,j} \vert i > < j \vert
\label{Hgene}
\end{eqnarray}
where $\epsilon_i$ is the random on-site energy on site $i$
and where $V_{i,j}$ is the hopping between the sites $i$ and $j$
(usually taken to be unity $V=1$ between nearest neighbors).

In dimension $d=1$, 
the transfer matrix formulation of the Schr\"odinger 
equation yields a log-normal distribution 
for the Landauer transmission $T_L$ \cite{anderson_fisher,luck}
\begin{eqnarray}
\ln T_L^{(d=1)} \oppropto_{L \to \infty} - \frac{L}{\xi_{loc}} + L^{1/2} u
\label{trans1d}
\end{eqnarray}
The leading non-random term is extensive in $L$ and involves
the localization length $\xi_{loc}$.
The subleading random term is of order $L^{1/2}$,
and the random variable $u$ of order $O(1)$ is Gaussian 
distributed as a consequence of the Central
Limit theorem.
Although it has been very often assumed and written
that this log-normal distribution 
persists in the localized phase in dimension $d=2,3$, 
theoretical arguments  \cite{NSS,medina}
and recent numerical calculations \cite{prior,rganderson}
are in favor of the following scaling
 form for the logarithm of the transmission
\begin{eqnarray}
\ln T_L^{(d)} \oppropto_{L \to \infty} -  \frac{L}{\xi_{loc}} + L^{\omega(d)} u
\label{transd}
\end{eqnarray}
where the exponent $\omega(d)$ depends on the dimension $d$ 
and coincides with the droplet exponent 
characterizing the strong disorder
phase of the directed polymer in a
random medium of dimension $1+D$ with $D=d-1$ (see the reminder on Directed Polymers 
in Appendix \ref{app_DP}). The probability distribution of the rescaled variable $u$
is not Gaussian but is determined by the directed polymer
universality class (see \cite{prior} where its distribution in $d=2$
is shown to coincide with the exactly known Tracy-Widom distribution
for the directed polymer in $1+1$).

The arguments in favor of the same universality class
can be decomposed in two steps \cite{NSS,medina,prior} :

(i) in the localized phase of Anderson localization in dimension $d$,
the transmission decays exponentially with the length, and thus 
{\it directed paths } completely dominate asymptotically over non-directed
paths.

(ii) these directed paths of the Anderson model
have weights that are random both in magnitude and sign,
but it turns out that the directed polymer model which is usually defined
with random positive weights (Boltzmann weights) keeps the same 
exponents in the presence of complex weights  
(see section 6.3 of the review \cite{Hal_Zha}).

Note that the Directed Polymer model fixes the fluctuation exponent $\omega$,
the rescaled distribution of the random variable $u$,
but does not give information on the divergence of the localization length $\xi_{loc} $
near the Anderson transition.
We expect that the same conclusions hold for the random transverse fields Ising model
(see section \ref{extension}).

% \newpage

\end{document}